%% file: ISIT_camera_ready.tex
\documentclass[conference]{IEEEtran}
\IEEEoverridecommandlockouts

\input{packages}
\input{commands}
\input{custom_formatting}

\input{plots_and_diagrams}
\input{custom_math_symbols}
\input{acronyms}

\begin{document}
\title{Pruning Neural Belief Propagation Decoders\\
\thanks{
    This work was presented at the \emph{IEEE International Symposium on Information Theory (ISIT) 2020}.
    This work was partially funded by the EU Horizon 2020 research and innovation programme under the Marie Sk\l{}odowska-Curie grant agreements no. 676448 and no. 749798 and by the Swedish Research Council under grant 2016-04253.
    Parts of the simulations were performed on resources at C3SE provided by the Swedish National Infrastructure for Computing (SNIC).
}
}

\author{
  \IEEEauthorblockN{
    Andreas Buchberger\IEEEauthorrefmark{1},
    Christian H\"ager\IEEEauthorrefmark{1},
    Henry D. Pfister\IEEEauthorrefmark{2},
    Laurent Schmalen\IEEEauthorrefmark{3}, and
    Alexandre Graell i Amat\IEEEauthorrefmark{1}
  }
  \IEEEauthorblockA{
    \IEEEauthorrefmark{1}Department of Electrical Engineering, Chalmers University of Technology, SE--412 96 Gothenburg, Sweden
  }
	\IEEEauthorblockA{
    \IEEEauthorrefmark{2}Department of Electrical and Computer Engineering, Duke University, Durham, North Carolina, USA
  }
	\IEEEauthorblockA{
    \IEEEauthorrefmark{3}Communications Engineering Lab, Karlsruhe Institute of Technology (KIT), 76131 Karlsruhe, Germany}
  }

\maketitle
\acresetall

\begin{abstract}
  We consider near \ac{ML} decoding of short linear block codes based on neural belief propagation (BP) decoding recently introduced by Nachmani \emph{et al.}. While this method significantly outperforms conventional BP decoding, the underlying parity-check matrix may still limit the overall performance. In this paper, we introduce a method to tailor an overcomplete parity-check matrix to (neural) BP decoding using machine learning. We consider the weights in the Tanner graph as an indication of the importance of the connected \acp{CN} to decoding and use them to prune unimportant \acp{CN}. As the pruning is not tied over  iterations, the final decoder uses a different parity-check matrix in each iteration. For \acl{RM} and short \acl{LDPC} codes, we achieve performance within \(\SI{0.27}{\decibel}\) and \(\SI{1.5}{\decibel}\) of the \ac{ML} performance while reducing the complexity of the decoder.
\end{abstract}
\acresetall

\section{Introduction}
For short code lengths, algebraic codes such as \ac{BCH} codes and \ac{RM} codes show excellent performance under \ac{ML} decoding. However, achieving near-\ac{ML} performance is computationally complex. A popular low-complexity decoding algorithm for block codes is \ac{BP} decoding. For \ac{LDPC} codes with sufficiently sparse parity-check matrices, \ac{BP} decoding provides near-optimal performance. However, for linear block codes with  dense parity-check matrices such as \ac{BCH} and \ac{RM} codes, the performance is not competitive. One reason for this is that the performance of \ac{BP} decoding can be significantly limited by  many short cycles in the graph.

Fueled by the advances in the field of deep learning, deep neural networks have also gained interest in the coding community \cite{Nachmani2016,Nachmani2018, Lian2019,Gruber2017}. In \cite{Nachmani2016,Nachmani2018}, \Ac{BP} decoding is formulated as a deep neural network. Instead of iterating between \acp{CN} and \acp{VN}, the messages are passed through unrolled iterations in a feed-forward fashion. In each iteration, the \acp{VN} and \acp{CN} are now referred to as \emph{\ac{VN} layers} and \emph{\ac{CN} layers}, respectively. Additionally, weights can be introduced at the edges which then are optimized using stochastic gradient descent (and variants thereof). This decoding method is commonly referred to as neural \ac{BP} and can be seen as a version of \ac{WBP} where each edge has a different weight. The idea is that the weights in the Tanner graph can account for short cycles and scale messages accordingly. In \cite{Lian2019}, the effects of coupling the weights over the iterations or over the nodes to reduce complexity was explored.

While \ac{WBP} decoding improves upon conventional \ac{BP} decoding, its performance is still limited by the underlying parity-check matrix. As the choice of the parity-check matrix is not unique, different choices of parity-check matrices may yield different performance. This fact has been exploited by using redundant parity-check matrices \cite{Kothiyal2005,Jiang2006,Halford2006,Hehn2010,Santi2018}. Kothiyal \emph{et al.} combined reliability-based decoding (e.g., ordered-statistics decoding) and \ac{BP} decoding in a scheme where the parity-check matrix is adapted to the outcome of the reliability-based decoding at the expense of high complexity\cite{Kothiyal2005}. In \cite{Jiang2006}, Reed-Solomon codes are decoded iteratively by adapting the parity-check matrix in each iteration while ensuring a practical complexity. In \cite{Halford2006}, a single Tanner graph is constructed from multiple parity-check matrices based on the permutation group of the code. In \cite{Hehn2010}, \ac{MBBP} decoding is introduced where \ac{BP} decoding is performed on multiple parity-check matrices in parallel. For \ac{RM} codes, a decoder adapting the parity-check matrix depending on the location of the most reliable bits was introduced in \cite{Santi2018}.

In this paper, we introduce a pruning-based approach to selecting the best parity-check equations for each iteration of the \ac{BP} decoder for short linear block codes.
Our pruning-based approach starts with a large overcomplete parity-check matrix under \ac{WBP} decoding. Considering the weights in the Tanner graph, the magnitude of the weights gives an indication of the importance of the edge in the decoding process. A magnitude close to zero indicate that the edge has low importance. By tying the weights for each \ac{CN}, i.e., enforcing that the weights of all incoming edges to a single \ac{CN} are equal,  the weights can be interpreted as an indication of the importance of the \ac{CN} in the decoding process. \acp{CN} with connected low-weight edges do not play an important role in the decoding process and can be removed.
We use this magnitude-based pruning approach to reduce the complexity of the decoder by removing \acp{CN} from the Tanner graph. By allowing pruning of different \acp{CN} in each iteration, the optimization results in a different parity-check equation for each iteration. The optimized parity-check matrices can be used to design decoders of different complexity, depending on the level of pruning. The weights in the corresponding Tanner graph can be untied, leading to the largest complexity. The weights obtained during the optimization process can be directly used. Alternatively, to achieve the lowest complexity, no weights at all may be used. For \ac{RM} and short \ac{LDPC} codes, we show that
this optimization improves performance over conventional \ac{BP} for the same complexity. In particular, the \ac{RM}(\(3,7\)) code performs within \(\SI{0.27}{\decibel}\) of the \ac{ML} performance. Also, a rate \(0.5\)-\ac{LDPC} code of length \(128\) performs within \(\SI{1.5}{\decibel}\) of the \ac{ML} performance, giving an improvement of \(\SI{0.5}{\decibel}\) over conventional \ac{BP}.

\section{Preliminaries}
Consider a linear block code \(\mathcal{C}\) of length \(n\) and dimension \(k\) with parity-check matrix \(\bm{H}\) of size \(m \times n\) where \(m \ge n - k\). If \(m > n - k\), we refer to the parity-check matrix as overcomplete and denote it as \(\bm{H}_\mathsf{oc}\). We denote the corresponding Tanner graph as \(\mathcal{G} = (\mathcal{V}_\mathsf{v}, \mathcal{V}_\mathsf{c}, \mathcal{E})\), consisting of a set of \acp{CN} \(\mathcal{V}_\mathsf{c}\), \(|\mathcal{V}_\mathsf{c}| = m\), a set of \acp{VN} \(\mathcal{V}_\mathsf{v}\), \(|\mathcal{V}_\mathsf{v}| = n\), and a set of edges \(\mathcal{E}\) connecting \acp{CN} with \acp{VN}.

For each \ac{VN} \(v\in \mathcal{V}_\mathsf{v}\) we define its neighborhood
\begin{align}
  \mathcal{N}(v) &\triangleq \left\{c\in\mathcal{V}_\mathsf{c}:(v,c)\in \mathcal{E}\right\},
\end{align}
i.e., the set of all \acp{CN} connected to \ac{VN} \(v\). Equivalently, we define the neighborhood of a \ac{CN} \(c\in \mathcal{V}_\mathsf{c}\) as
\begin{align}
  \mathcal{N}(c) &\triangleq \left\{v\in\mathcal{V}_\mathsf{v}:(v,c)\in \mathcal{E}\right\}.
\end{align}
Let \(\lambda_{v\rightarrow c}^{(\ell)}\) and \(\lambda_{c\rightarrow v}^{(\ell)}\) be the message passed from \ac{VN} \(v\) to \ac{CN} \(c\) and the message passed from \ac{CN} \(c\) to \ac{VN} \(v\), respectively, in the \(\ell\)-th iteration. The \ac{VN} and \ac{CN} updates are
\begin{align}
  \lambda_{v\rightarrow c}^{(\ell)} &=  \lambda_{\mathsf{ch},v} + \sum_{\tilde{c}\in \mathcal{N}(v)\backslash c}  \lambda_{\tilde{c} \rightarrow v}^{(\ell)}
  \label{eq:vn_update}
\end{align}
and
\begin{align}
  \lambda_{c\rightarrow v}^{(\ell)} &= 2\tanh^{-1}\left(\prod_{\tilde{v}\in \mathcal{N}(c)\backslash v}  \tanh\left(\frac{1}{2}\lambda_{\tilde{v} \rightarrow c}^{(\ell)}\right)\right),
  \label{eq:cn_update}
\end{align}
respectively, where \(\lambda_{\mathsf{ch},v}\) is the \ac{LLR} of the channel output. For bipolar transmission over the additive white Gaussian noise channel, it follows that
\begin{align}
 \lambda_{\mathsf{ch},v} &\triangleq \ln \frac{p_{Y|B}(y_v|b_v=0)}{p_{Y|B}(y_v|b_v=1)} \overset{}{=}  \frac{2y_v}{\sigma^2}
\end{align}
where \(y_v\) is the channel output, \(b_v\) is the transmitted bit, and \(\sigma^2\) is the variance of the noise.
The \ac{VN} output \ac{LLR} in the \(\ell\)-th iteration is
\begin{align}
  \lambda_{v}^{(\ell)} &=  \lambda_{\mathsf{ch},v} + \sum_{\tilde{c}\in \mathcal{N}(v)}  \lambda_{\tilde{c} \rightarrow v}^{(\ell)}.
  \label{eq:vn_marginalization}
\end{align}

\subsection{Weighted Belief Propagation}
One way to counteract the effect of short cycles on the \ac{BP} decoding performance is to introduce weights for each edge in the Tanner graph \cite{Nachmani2016,Nachmani2018} which is referred to as neural \ac{BP} and can be seen as version of \ac{WBP} where each edge has an individual weight. For \ac{WBP}, the update rules \internalEq{eq:vn_update} and \internalEq{eq:cn_update}  modify to
\begin{align}
  \lambda_{v\rightarrow c}^{(\ell)} &=  w_{v}^{(\ell)}\lambda_{\mathsf{ch},v}+ w_{v\rightarrow c}^{(\ell)}\sum_{\tilde{c}\in \mathcal{N}(v)\backslash c}  \lambda_{\tilde{c} \rightarrow v}^{(\ell)}.
  \label{eq:vn_update_wbp}
\end{align}
and
\begin{align}
  \lambda_{c\rightarrow v}^{(\ell)} &= 2 w_{c\rightarrow v}^{(\ell)}\tanh^{-1}\left(\prod_{\tilde{v}\in \mathcal{N}(c)\backslash v}  \tanh\left(\frac{1}{2}\lambda_{\tilde{v} \rightarrow c}^{(\ell)}\right)\right)
  \label{eq:cn_update_wbp}
\end{align}
where \( w_{v}^{(\ell)}\), \(w_{v\rightarrow c}^{(\ell)}\), and \(w_{c\rightarrow v}^{(\ell)}\), are the channel, \ac{VN}, and \ac{CN} weights, respectively.
The \ac{VN} output \ac{LLR} in the \(\ell\)-th iteration is
\begin{align}
  \lambda_{v}^{(\ell)} &=  w_{v}^{(\ell)}\lambda_{\mathsf{ch},v} + w_{v\rightarrow c}^{(\ell)}\sum_{\tilde{c}\in \mathcal{N}(v)}  \lambda_{\tilde{c} \rightarrow v}^{(\ell)}.
  \label{eq:vn_marginlaization_wbp}
\end{align}
Update rules \internalEq{eq:vn_update_wbp} and \internalEq{eq:cn_update_wbp} describe \ac{WBP} when the weights are untied over all nodes as well as over all iterations. In order to reduce complexity, the weights can also be tied. In \cite{Lian2019}, tying weights temporally, i.e., over  iterations, and tying the weights spatially, i.e., within one node layer, was explored.

Here, we consider the case where the weights in each \ac{CN} are tied, i.e., \(w_{c}^{(\ell)} = w_{c\rightarrow v}^{(\ell)}\) for all \(v\in\mathcal{N}(c)\). Hence, the \ac{CN} update results to
\begin{align}
  \lambda_{c\rightarrow v}^{(\ell)} &= 2w_c^{(\ell)}\tanh^{-1}\left(\prod_{\tilde{v}\in \mathcal{N}(c)\backslash v}  \tanh\left(\frac{1}{2} \lambda_{\tilde{v} \rightarrow c}^{(\ell)}\right)\right).
  \label{eq:cn_update_pp}
\end{align}
Note that setting all the weights of a \ac{WBP} decoder to one results in conventional \ac{BP} decoding.

\subsection{Optimization of the Weights}
The decoding process can be seen as a classification task where the channel output is mapped to a valid codeword. This task consists of \(2^k\) classes, one for each codeword. Training such a classification task is unfeasible as the resulting decoder generally generalizes poorly to classes not contained in the training data \cite{Gruber2017}. Instead, the task can be reduced to binary classification for each of the \(n\) bits. As a loss function, the bitwise cross-entropy between the transmitted codeword and the \ac{VN} output \ac{LLR} of the final \ac{VN} layer was used in \cite{Nachmani2016,Nachmani2018}.
The optimization behavior can be improved by using a multiloss, where the overall loss is the average  bitwise cross-entropy between the transmitted codeword and the \ac{VN} output \ac{LLR} of each \ac{VN} layer.

In \cite{Lian2019}, it was observed that the binary-cross entropy does not perform well for large, overcomplete parity-check matrices. Wrongly decoded bits with large \acp{LLR} result in large cross-entropy losses and cause the training to converge slowly. As an alternative, the loss function
\begin{align}
\Gamma &=  \frac{1}{n}\sum_{v=1}^n (1-o_v)^{x_v} o_v^{1-x_v}
\label{eq:soft_ber}
\end{align}
was proposed, where \(o_v^{(\ell)}\) is the estimate of the probability that the \(v\)-th bit in the \(\ell\)-th iteration is one, i.e., \(o_v^{(\ell)} = \sigma (\lambda_v^{(\ell)}) = 1/(1+\mathrm{exp}(-\lambda_v^{(\ell)}))\). Since substituting hard-decision values for \(o_v^{(\ell)}\) results in the bit-error rate, \internalEq{eq:soft_ber} is referred to as soft bit-error rate.
Combining the soft bit-error rate and a multiloss results to
\begin{align}
\tilde{\Gamma} &= \frac{1}{\sum \eta^{L-\ell}}\sum_{t=1}^L \eta^{L-\ell} \frac{1}{n}\sum_{v=1}^n \left(1-o_v^{(\ell)}\right)^{x_v} \left(o_v^{(\ell)}\right)^{1-x_v}\\
&\overset{(a)}{=}\frac{1}{\sum \eta^{L-\ell}}\sum_{\ell=1}^L \eta^{L-\ell} \frac{1}{n}\sum_{v=1}^n  o_v^{(\ell)}
\label{eq:loss}
\end{align}
where \(\eta\) determines the contribution of intermediate layers to the overall loss and is decreased during the training, such that in the final phase of the training only the last layer contributes to the loss \cite{Lian2019}. Step \((a)\) follows from the fact that since the channel and the decoder are symmetric, the all-zero codeword can be used for training.

\section{Optimizing the Parity-Check Matrix}
While \ac{WBP} decoding as described in the previous section improves upon conventional \ac{BP} decoding, its performance is quite dependent on the choice of the parity-check matrix.
Here, we propose a pruning-based approach to select the relevant parity-checks from a large, overcomplete parity-check matrix.
For this, we consider a modified \ac{WBP} where the weights are tied at the \acp{CN}, i.e., all messages at a single \ac{CN} are weighted by the same weight \(w_c^{(\ell)}\) as in \internalEq{eq:cn_update_pp}. The \ac{VN} update \internalEq{eq:vn_update_wbp} remains unchanged.
The magnitude of the weights \(w_c^{(\ell)}\) can now be interpreted as a measure of how much the \ac{CN} contributes to decoding. A large magnitude indicates high importance whereas a magnitude of zero indicates that the \ac{CN} is irrelevant to the decoding process.

\subsection{Training Procedure}
\begin{figure}
     \centering
     \includegraphics{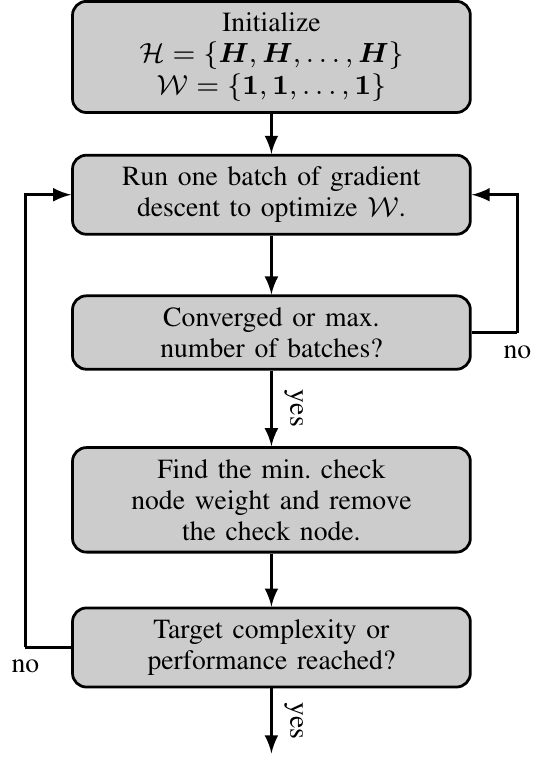}
     \caption{Flowchart of the training process.}
     \label{fig:training}
 \end{figure}

Let \(\mathcal{H} = \{\bm{H}_1, \ldots, \bm{H}_L\}\) be a set of parity-check matrices where \(\bm{H}_\ell\) is the parity-check matrix used for decoding in the \(\ell\)-th iteration. Equivalently, we define a set of weights \(\mathcal{W}\).
The set of parity-check matrices is initialized with the same large overcomplete matrix \(\bm{H}_\mathsf{oc}\) for each iteration, i.e., \(\bm{H}_\ell = \bm{H}_\mathsf{oc}\), \(\ell=1,\ldots, L\). All weights are initialized to one, i.e., we start with conventional \ac{BP}. The weights in \(\mathcal{W}\) are then optimized using the Adam optimizer \cite{Kingma2014} within the Tensorflow programming framework \cite{tensorflow2015}.  After the optimization has converged, we find the index and the iteration of the lowest \ac{CN} weight \(w_c^{(\ell)}\) and set it to zero, i.e., we prune the corresponding parity-check equation from \(\mathcal{W}\). As this may change the optimal value for the remaining weights, we rerun the training.
We iterate between retraining and pruning \acp{CN} until we either reach a desired number of parity-check equations or until the loss starts diverging.
The result of the optimization is a set of parity-check equations \(\mathcal{H}_\mathsf{opt} = \left\{\bm{H}_{\ell,\mathsf{opt}}\right\}\) with \(\ell=1,\ldots, L\) and optimized weights \(\mathcal{W}_\mathsf{opt}\). The training process is  illustrated in the flowchart of  \internalFig{fig:training}.

\subsection{Complexity Discussion}
  In the \ac{BP} decoder, the evaluation of the  \(\tanh\) and inverse \(\tanh\) functions is the operation of highest computational complexity. It is natural to use the number of these evaluations as a measure of the complexity of the decoder. As both functions are evaluated in the \acp{CN}, it is equivalent to use the number of \acp{CN} and hence the number of parity-check equations, i.e., rows in the parity-check matrix.
  The required memory is related to the parity-check matrix itself and the number of weights. Since the weights are real numbers as opposed to binary values for the edges, we quantify memory requirements with the number of weights.

  With this, we define three decoders of different complexity.
  \begin{itemize}
    \item Decoder \(\mathcal{D}_1\): It uses the result from the optimization directly, i.e.,   \(\mathcal{H}_\mathsf{opt}\) and \(\mathcal{W}_\mathsf{opt}\). Hence, it uses \internalEq{eq:vn_update_wbp} and \internalEq{eq:cn_update_pp} as updates in the respective nodes.
    \item Decoder \(\mathcal{D}_2\): It uses the optimized set of parity-check matrices, i.e.,  \(\mathcal{H}_\mathsf{opt}\), but sets all weights to one, i.e., neglects  \(\mathcal{W}_\mathsf{opt}\). It uses \internalEq{eq:vn_update} and \internalEq{eq:cn_update} as updates in the respective nodes.
    \item Decoder \(\mathcal{D}_3\):  It uses the optimized set of parity-check matrices, i.e.,  \(\mathcal{H}_\mathsf{opt}\), and additionally untied optimized weights over all iterations and edges as in \cite{Nachmani2016}. Hence, the updates rules become \internalEq{eq:vn_update_wbp} and \internalEq{eq:cn_update_wbp}   in the respective nodes. It is important to note that to obtain the untied weights, an extra training step with untied weights is required as previously we only considered tied weights in the \acp{CN}.
  \end{itemize}
Concerning the  decoders, all three  have similar computational complexity as they operate on the same set of parity-check matrices. However, they differ in the required memory. Decoder \(\mathcal{D}_3\) needs to store the most weights, i.e., one weight per edge, whereas \(\mathcal{D}_2\) does not need to store any weights. Decoder \(\mathcal{D}_1\) only needs to store one weight per \ac{CN} and hence is of lower complexity than  \(\mathcal{D}_3\) but higher complexity than  \(\mathcal{D}_2\).
\vspace{-2pt}
\section{Numerical Results}
\vspace{-2pt}
We numerically evaluate the performance of the proposed parity-check matrix optimization for \ac{RM} codes and a short \ac{LDPC} code. As a benchmark we consider \ac{ML} decoding.
\vspace{-2pt}
\subsection{The \Acl{RM} Code RM\((2,5)\)}
\vspace{-2pt}
\begin{figure}
    \centering
    \includegraphics{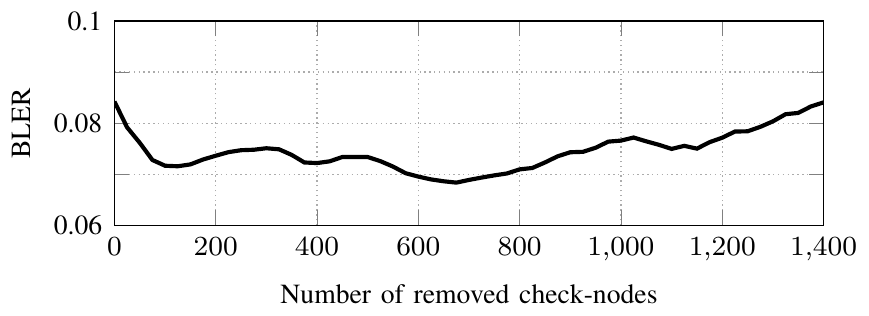}
    \caption{BLER at \(\SI{3}{\decibel}\) during the training process for the \ac{RM}\((2,5)\) code.}
    \label{fig:pruning_rm_2_5}
\end{figure}
\begin{figure}
    \centering
    \includegraphics{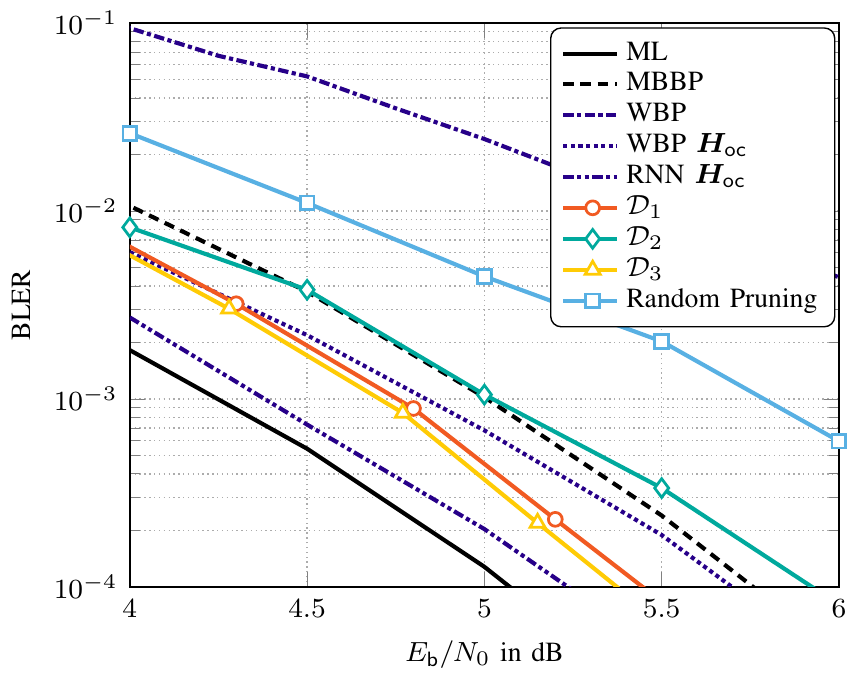}
    \caption{BLER results for the RM\((2, 5)\) code.}
    \label{fig:bler_rm_2_5}
\end{figure}

The RM\((2, 5)\) code has \(620\) parity-check equations of minimum weight that are used to initialize the optimization. We fix the number of iterations to six. Hence,  \(620\) \acp{CN} need to be evaluated per iteration which leads to a total of \(3720\) \acp{CN} that need to be evaluated.
The optimization is stopped when the loss starts to increase. In total, \(\SI{31}{\percent}\) of the parity-check equations remain. In \internalFig{fig:pruning_rm_2_5}, we depict the \ac{BLER} as a function of removed \acp{CN}. In \internalFig{fig:bler_rm_2_5}, we plot the \ac{BLER} as a function of $E_\mathsf{b}/N_0$.
Decoder \(\mathcal{D}_1\) performs  within \(\SI{0.38}{\decibel}\) of the \ac{ML} decoder at a \ac{BLER} of \(10^{-4}\). Removing the weights from the optimized parity-check matrix (\(\mathcal{D}_2\)), results in a penalty of \(\SI{0.48}{\decibel}\).  Untying the weights in the \acp{CN} (\(\mathcal{D}_3\)) results in an additional gain of \(\SI{0.047}{\decibel}\) with respect to \(\mathcal{D}_2\).

Both \(\mathcal{D}_1\) and \(\mathcal{D}_3\), requiring \(1170\) \acp{CN}, outperform \ac{WBP} \cite{Nachmani2016} with \(\bm{H}_\mathsf{oc}\) containing the \(620\) parity-check equations of minimum weight and hence \(3720\) \acp{CN}, as well as  \ac{MBBP} \cite{Hehn2010} with \(15\) randomly chosen parity-check matrices with \(1440\) \acp{CN}. Only  \(\mathcal{D}_2\) performs slightly worse, but at the same time requires only \(1170\) \acp{CN} and no weights. A recurrent neural network (RNN)-based decoder as introduced in \cite{Nachmani2018} using \(\bm{H}_\mathsf{oc}\) slightly outperforms \(\mathcal{D}_1\) and \(\mathcal{D}_2\) at the cost of increased complexity by about three times. \ac{WBP} \cite{Nachmani2016} with a standard parity-check matrix containing \(16\) \acp{CN} is clearly not competitive.
The decoding complexity of the decoders in  \internalFig{fig:bler_rm_2_5} is reported in \internalTab{tab:complexity}.

To verify the effectiveness of our pruning strategy for \acp{CN}, we also consider the scenario where we randomly prune \acp{CN}. As it can be observed in \internalFig{fig:bler_rm_2_5}, this approach is clearly not competitive.

\begin{figure}[!t]
    \centering
    \includegraphics{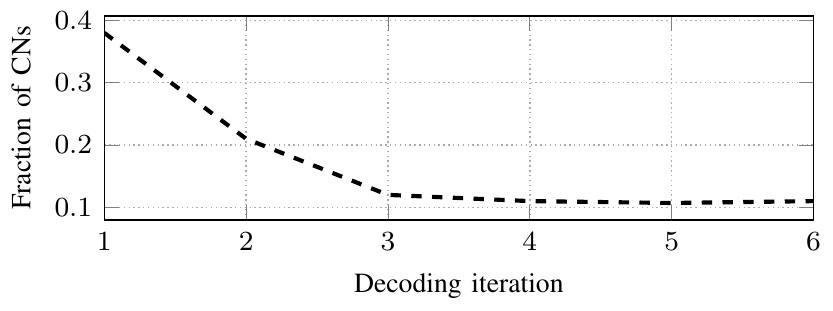}
    \caption{Fraction of the remaining \acp{CN} after pruning used for decoding in the respective iteration for the RM\((2,5)\) code.}
    \label{fig:rm_2_5_cn_fraction}
\end{figure}

To investigate the behavior of the pruning, we are interested how many \acp{CN} are pruned in each \ac{BP} iteration, or equivalently, how many \acp{CN} remain. To this end, we plot the fraction of all remaining \acp{CN} per iteration in \internalFig{fig:rm_2_5_cn_fraction}. We observe that in the first \ac{BP} iteration, about \(\SI{40}{\percent}\) of all remaining \acp{CN} are used for decoding. In later \ac{BP} iterations, the number of \acp{CN} decreases significantly. This observation furthermore justifies the use of a low number of iterations.

\vspace{-5pt}

\ctable[
caption = {Complexity of the decoders for different codes.},
label   = tab:complexity,
pos     = tb,
doinside = \small,
]{lcr}{
}{
\FL     & \# of \ac{CN} evaluations  \LL
WBP, RM\((2, 5)\)  & \(16\cdot 6 = 96\)\NN
WBP, RNN \(\bm{H}_\mathsf{oc}\), RM\((2, 5)\)  & \(620\cdot 6 = 3720\)\NN
 \ac{MBBP} RM\((2, 5)\) & \(15\cdot 6\cdot 16 = 1440\) \NN
\(\mathcal{D}_1\), \(\mathcal{D}_2\), \(\mathcal{D}_3\),  RM\((2, 5)\)  & \(620 \cdot 6\cdot 0.31 = 1170\)\NN
Random, RM\((2, 5)\)  &   \(620 \cdot 6\cdot 0.31 = 1170\)\LL
WBP, RNN \(\bm{H}_\mathsf{oc}\), RM\((3, 7)\)  & \(94488\cdot 6 = 566928\)\NN
\ac{MBBP} RM\((3, 7)\)& \(60\cdot 6\cdot 64 = 23440\) \NN
\(\mathcal{D}_1\), \(\mathcal{D}_2\), \(\mathcal{D}_3\), RM\((3, 7)\)  & \(94488 \cdot 6\cdot 0.03 = 19842\)\NN
\(\mathcal{\tilde{D}}_1\), RM\((3, 7)\)  & \(9448 \cdot 6\cdot 0.35 = 19842\)\LL
BP, CCSDS, 25 iterations  & \(64\cdot 25 = 1600\)\NN
BP, CCSDS, 100 iterations  & \(64\cdot 100 = 6400\)\NN
WBP, CCSDS  & \(64\cdot 6 = 96\)\NN
\(\mathcal{D}_1\), \(\mathcal{D}_3\) CCSDS  & \(10000 \cdot 6\cdot 0.027 = 1600\)\LL
\vspace{-13pt}
}

\subsection{The \Acl{RM} Code RM\((3,7)\)}
\vspace{-2pt}

The \ac{RM}\((3,7)\) code has \(94488\) parity-check equations of minimum weight. In the initial training phases, optimizing the weights converges very slowly and removing \acp{CN} is done in an almost random fashion. To speed up the training for decoders \(\mathcal{D}_1\), \(\mathcal{D}_2\), and \(\mathcal{D}_3\), we randomly select \(70000\) parity-check equations and use them as the  overcomplete parity-check matrix \(\bm{H}_\mathsf{oc}\). Decoder \(\mathcal{\tilde{D}}_1\) uses only a small, random subset, namely \(9448\),  of all parity-check equations of minimum weight as the overcomplete parity-check matrix \(\bm{H}_\mathsf{oc}\). For all four decoders, we consider six iterations.

\begin{figure}
    \centering
    \includegraphics{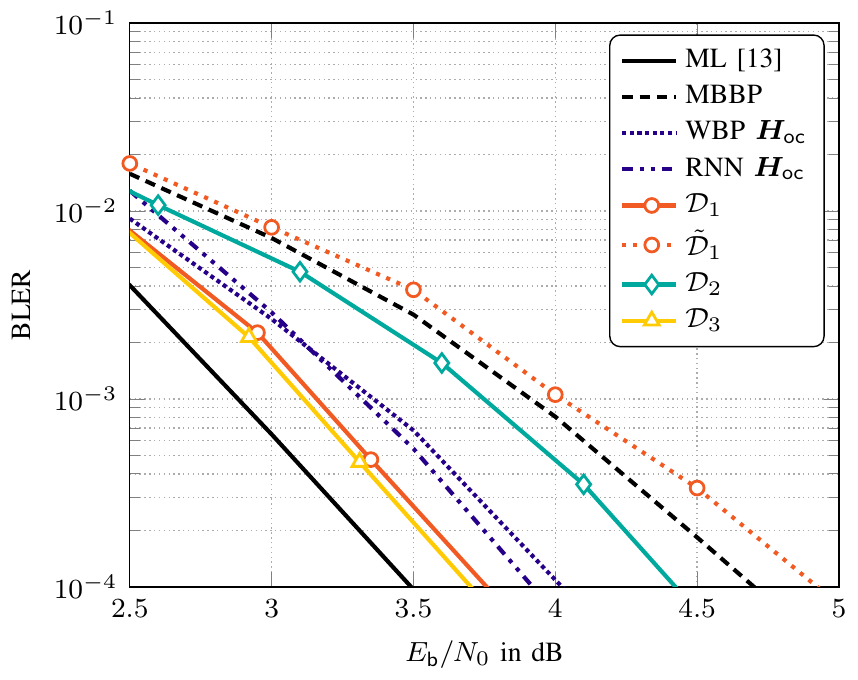}
    \caption{BLER results for the RM\((3, 7)\) code.}
    \label{fig:bler_rm_3_7}
\end{figure}

In \internalFig{fig:bler_rm_3_7}, we plot the \ac{BLER}. Decoder \(\mathcal{D}_1\) performs  within \(\SI{0.27}{\decibel}\) of the \ac{ML} decoder. Removing the weights results in a degradation of \(\SI{0.47}{\decibel}\) for decoder \(\mathcal{D}_2\) with respect to \(\mathcal{D}_1\). On the other hand, untying the weights results in a gain of \(\SI{0.02}{\decibel}\) for decoder \(\mathcal{D}_3\). Decoders \(\mathcal{D}_1\), \(\mathcal{D}_2\), and \(\mathcal{D}_3\), all require \(19842\) \acp{CN} and outperform  \ac{MBBP} with \(60\) randomly chosen parity-check matrices, i.e., \(23440\) \acp{CN}, and \ac{WBP} as in \cite{Nachmani2016} with \(\bm{H}_\mathsf{oc}\) and the RNN-based decoder \cite{Nachmani2018} with \(\bm{H}_\mathsf{oc}\) while having lower complexity. As for the \ac{RM}\((2,5)\) code, \ac{WBP} over the standard parity-check matrix with \(64\) \acp{CN} is not competitive (curve omitted for better readability). The complexities are reported in \internalTab{tab:complexity}.

Decoder \(\mathcal{\tilde{D}}_1\) demonstrates the effect of using only a small subset of all parity-check equations of minimum weight as the  overcomplete parity-check matrix \(\bm{H}_\mathsf{oc}\). In this case, only \(9448\) randomly chosen parity-check equations were used initially and the decoder is pruned to the same complexity as \(\mathcal{D}_1\). This essentially corresponds to randomly pruning \acp{CN} and results in the same performance degradation as for the \ac{RM}\((2,5)\) code.

\vspace{-4pt}
\subsection{Low-Density Parity-Check Code}
We consider the CCSDS \ac{LDPC} code of length \(128\) and rate \(0.5\) as defined in \cite{ccsds}. It has a \ac{CN} degree of \(8\) with half the \acp{VN} having degree \(3\) and half having degree \(5\). The code has a minimum distance \(14\). For decoding, \(25\) iterations are used. Hence, a total of \(1600\) \ac{CN} updates are required.

For the overcomplete matrix, we start with \(10000\) randomly chosen parity-check equations of low or minimum weight. Then, we prune the decoder to the same complexity as conventional \ac{BP} decoding, i.e., we allow \(1600\) \ac{CN} updates. The number of iterations is set to six. Both decoders \(\mathcal{D}_1\) and \(\mathcal{D}_3\) outperform conventional \ac{BP} by approximately \(\SI{0.6}{\decibel}\).
Allowing \(100\) iterations for conventional \ac{BP} shows that the gain of decoders \(\mathcal{D}_1\) and \(\mathcal{D}_3\) decreases to \(\SI{0.2}{\decibel}\). However, conventional \ac{BP} with \(100\) iterations requires \(6400\) \acp{CN} and therefore has a higher complexity than decoders \(\mathcal{D}_1\) and \(\mathcal{D}_3\). \ac{WBP} \cite{Nachmani2016} with \(64\) \acp{CN} is again not competitive. The complexities are reported in \internalTab{tab:complexity}.

\begin{figure}
    \centering
    \includegraphics{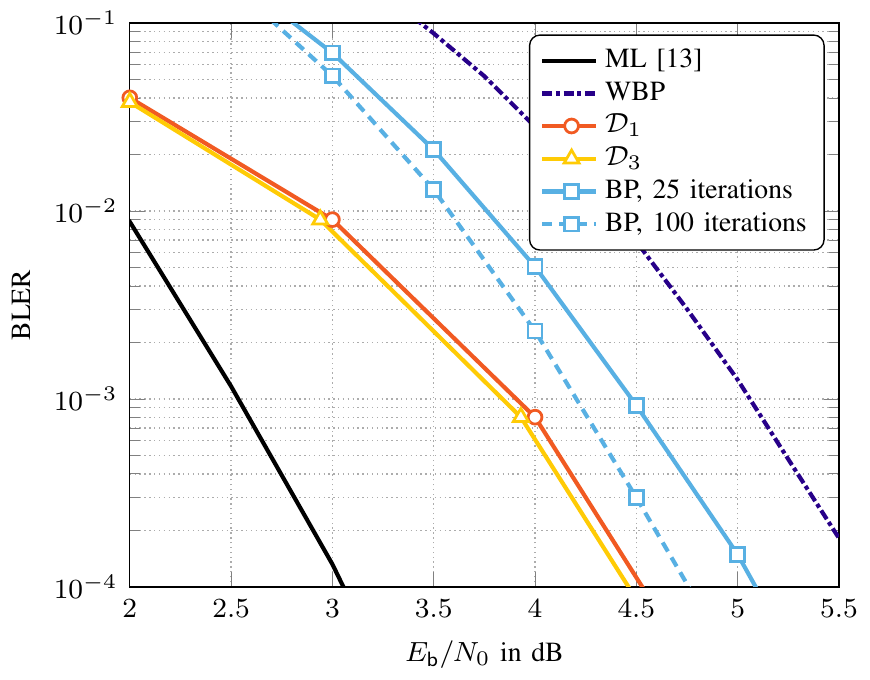}
    \caption{\Ac{BLER} results for the LDPC code.}
    \label{fig:bler_ldpc}
\end{figure}

\section{Conclusion}
We applied machine learning to optimize the parity-check matrix for conventional and weighted belief propagation decoding. To this end, we prune a large overcomplete parity-check matrix and allow it to consist of different parity-check equations in each iteration. We obtain significant performance gains while keeping the complexity practical.  For \ac{RM} and short, standardized \ac{LDPC} codes we demonstrated  a performance within up to \(\SI{0.27}{\decibel}\)  and \(\SI{1.5}{\decibel}\) of \ac{ML} decoding, respectively. In all scenarios, our approach outperforms conventional \ac{BP} while having equal complexity and \ac{MBBP} and the original neural \ac{BP} while even allowing lower complexity. Our approach can easily be applied to any other linear block code such as \ac{BCH} codes and similar gains over conventional \ac{BP} decoding are expected.
\bibliographystyle{IEEEtran}
\nocite{channelcodes}
\bibliography{IEEEabrv,CONFabrv,literature}

\end{document}

%% file: packages.tex
\usepackage[final]{listings}
\usepackage{acro}

\usepackage{xifthen}

\usepackage{cite}

\usepackage{nohyperref}

\usepackage{tikz}
\usepackage{pgfplots}
\usetikzlibrary{arrows, positioning, shapes, calc, spy}
\usepgfplotslibrary{fillbetween, groupplots}

\usepackage[cmex10]{amsmath}
\usepackage{amssymb,amsfonts,amsthm}

\usepackage{bm}

\usepackage{siunitx}

\usepackage{ctable}

\usepackage{xcolor}

\usepackage{algorithm,algorithmic}

\usepackage{ifthen}

\usepackage{marginnote}

%% file: commands.tex
\newcommand{\internalLink}[1]{\hyperref[#1]{\ref*{#1}}}
\newcommand{\internalLinkChapter}[1]{Chapter \hyperref[#1]{\ref*{#1}}}
\newcommand{\internalLinkSection}[1]{Section \hyperref[#1]{\ref*{#1}}}
\newcommand{\internalLinkAppendix}[1]{Appendix \hyperref[#1]{\ref*{#1}}}
\newcommand{\internalTab}[1]{Table\,\hyperref[#1]{\ref*{#1}}}
\newcommand{\internalFig}[1]{Fig.\,\hyperref[#1]{\ref*{#1}}}
\newcommand{\internalEq}[1]{(\hyperref[#1]{\ref*{#1}})}
\newcommand{\internalEqs}[2]{(\hyperref[#1]{\ref*{#1}})-(\hyperref[#2]{\ref*{#2}})}
\newcommand{\internalListing}[1]{Listing \hyperref[#1]{\ref*{#1}}}
\newcommand{\internalAlgorithm}[1]{Algorithm\,\hyperref[#1]{\ref*{#1}}}
\newcommand{\internalLemma}[1]{Lemma \hyperref[#1]{\ref*{#1}}}
\newcommand{\internalDefinition}[1]{Def. \hyperref[#1]{\ref*{#1}}}
\newcommand{\internalTheorem}[1]{Theorem \hyperref[#1]{\ref*{#1}}}

%% file: custom_formatting.tex
\let\originalleft\left
\let\originalright\right
\renewcommand{\left}{\mathopen{}\mathclose\bgroup\originalleft}
\renewcommand{\right}{\aftergroup\egroup\originalright}

%% file: plots_and_diagrams.tex
\definecolor{curve_color01}{RGB}{0, 0, 0}      %
\definecolor{curve_color04}{RGB}{88, 176, 227} 
\definecolor{curve_color03}{RGB}{0, 169, 157} 
\definecolor{curve_color02}{RGB}{ 39, 0, 137}   
\definecolor{curve_color05}{RGB}{255, 203, 5}   
\definecolor{curve_color07}{RGB}{92 75, 61}     
\definecolor{curve_color06}{RGB}{241, 90, 34}   
\definecolor{color_weak_black}{RGB}{20,20,20}

\definecolor{block_gray}{RGB}{204,204,204}
\definecolor{block_background_gray}{RGB}{240,240,240}
\definecolor{block_background_line_gray}{RGB}{175,175,175}
\definecolor{block_weak_gray}{RGB}{240,240,240}

\tikzstyle{line}=[draw, thick]
\tikzstyle{arrow}=[draw, -latex, thick, midway]
\tikzstyle{line}=[draw, -, thick, midway]

\tikzstyle{block} = [rectangle, draw, thick, fill=block_gray,
    text width=2.75em, text centered, rounded corners, minimum height=1.25em]
\tikzstyle{block_circle} = [circle, draw, thick, fill=block_gray, text centered]
\tikzstyle{channel} = [cloud, draw,thick, cloud puffs=12, cloud puff arc=120, aspect=2, fill=block_gray]

\tikzstyle{block_background} = [rectangle, draw, align=center, thick, draw=block_background_line_gray, fill=block_background_gray, rounded corners]

\pgfplotsset{
  legend style={font=\footnotesize},
  every tick label/.append style={font=\footnotesize},
  every axis label/.append style ={font=\footnotesize},
	legend image code/.code={
		\draw[mark repeat=2,mark phase=2]
			plot coordinates {
				(0cm,0cm)
				(0.3cm,0cm)        
				(0.55cm,0cm)         
			};%
	}
}

%% file: custom_math_symbols.tex
\DeclareSIUnit{\belmilliwatt}{Bm}
\DeclareSIUnit{\dBm}{\deci\belmilliwatt}
\DeclareSIUnit{\bit}{bit}
\DeclareSIUnit{\bits}{bits}

%% file: acronyms.tex
\DeclareAcronym{BCH}{
	short = BCH,
	long= Bose-Chaudhuri-Hocquenghem
}

\DeclareAcronym{BER}{
	short = BER,
	long= bit error rate
}

\DeclareAcronym{BLER}{
	short = BLER,
	long= block error rate
}

\DeclareAcronym{BP}{
	short = BP,
	long = belief propagation
}

\DeclareAcronym{CN}{
	short = CN,
	long = check node
}

\DeclareAcronym{LDPC}{
	short = LDPC,
	long = low-density parity-check
}

\DeclareAcronym{LLR}{
	short = LLR,
	long = log-likelihood ratio
}

\DeclareAcronym{MBBP}{
	short = MBBP,
	long = multiple-bases belief propagation
}

\DeclareAcronym{ML}{
	short = ML,
	long = maximum-likelihood
}

\DeclareAcronym{RM}{
	short = RM,
	long= Reed-Muller
}

\DeclareAcronym{VN}{
	short = VN,
	long = variable node
}

\DeclareAcronym{WBP}{
	short = WBP,
	long = weighted belief propagation
}